\documentstyle[12pt]{article}
\begin{document}

\begin{center}
{\it  Work presented at the Sixth 
International Conference on the Optics of
Excitons in Confined Systems, Ascona, Switzerland, Aug. 30-Sept. 2 ,1999.}
\end{center}
\vspace{1cm}
\centerline{\bf Decoherence effects on the generation of exciton
entangled states}
\centerline{\bf in coupled quantum dots}
\normalsize

\

\centerline{F.J.Rodr\'{\i}guez$^1$ (a), L.Quiroga$^2$(a) and N.F.Johnson$^3$ (b)}

\centerline{\em (a) Depto. de F\'{\i}sica, Universidad de los Andes,
A.A.
4976,
Santaf\'e de Bogot\'a, Colombia}

\centerline{\em (b) Physics Department, Clarendon Laboratory, Oxford
University,
Oxford OX1 3PU, U.K.}

\

\noindent
\begin{abstract}
\par We report on exciton-acoustic-phonon coupling effects on the
generation of exciton maximally entangled states in $N=2$ and $3$ quantum dot 
systems.
In particular, we address the question of the
combined effect of laser pulses, appropriate for generating Bell and
Greenberger-Horne-Zeilinger entangled states,
together with
decoherence mechanisms as provided by a phonon reservoir.
By solving numerically the master equation for the optically driven exciton-phonon
kinetics,
we show that the generation of maximally entangled exciton states is
preserved
over a reasonable parameter window.
\end{abstract}


\vspace{2cm}
\noindent
$^1$ frodrigu@uniandes.edu.co\\
$^2$ lquiroga@uniandes.edu.co\\
$^3$ n.johnson@physics.oxford.ac.uk\\
\newpage

%\section{Introduction}
\par Confined excitons together with ultrafast optical spectroscopy have
been
shown to be important elements for achieving coherent wavefunction
control
on the
nanometer and femtosecond scales in semiconductors \cite{bonadeo}.
Maximally
entangled states (MES), of
Bell-type for excitons in two coupled quantum
dots (QDs) and Greenberger-Horne-Zeilinger (GHZ) type for three coupled
QDs,
have been reported as
excellent candidates for achieving quantum entanglement in
solid-state based
devices \cite{qj}. 
However, the question arises as to how reliable the MES preparation
scheme of Ref. [2] will be when decoherence
mechanisms are taken into account during the generation step.
Exciton decoherence in semiconductor QDs is dominated by acoustic
phonon scattering at low temperatures \cite{taka}.
\par In this work we present results on the kinetics of
the generation of exciton MES in QDs,
taking into account an acoustic phonon dephasing mechanism. The
Hamiltonian
describing a system formed by $N$ QDs in the rotating wave approximation is
\begin{eqnarray}
H(t)=\Delta \omega J_z-V(J^2-J_z^2)-A(J^++J^-)+\sum_{\vec k}
{\omega_{\vec k}a_{\vec k}^{\dag} a_{\vec k}}+\sum_{\vec k}
{g_{\vec k}J_z(a_{\vec k}^{\dag}+a_{\vec k})}
\end{eqnarray}
where
$J_+=\sum_{n=1}^{N} {e_{n}^{\dag}h_{n}^{\dag}}$,
$J_-=\sum_{n=1}^{N}{h_{n}e_{n}}$ and
$J_z=\frac {1}{2} \sum_{n=1}^{N} {(e_{n}^{\dag}e_{n}-
h_{n}h_{n}^{\dag})}$
with $e_{n}^{\dag}$ ($h_n^{\dag}$) describing the electron (hole)
creation operator
in the $n$'th QD.
The collective operators describing the QD excitons, $J$-operators, satisfy 
the usual angular momentum commutation
relationships:
$[ J_+,J_-]=2J_z$,
$[ J_{\pm},J_z]=\mp J_{\pm}$. $\Delta \omega=\epsilon-\omega$ is the
resonance detuning, 
$\epsilon$ denotes the semiconductor energy gap, $\omega$ is the laser
central frequency, $V$ the
Forster term representing the Coulomb interdot interaction, $A$ the laser pulse 
amplitude and $a_{\vec k}^{\dag}$
($a_{\vec k}$)
the creation (annihilation) operator of the acoustic phonon with
wavevector ${\vec k}$. We put $\hbar=1$ throughout this paper.
We work within
corresponding optically active exciton states, $i.e.$
$J=1$ and
$J=3/2$ for two and three coupled quantum dots, respectively. Mixing
with
dark exciton states can be induced by exciting selectively a single QD
or by
a different coupling with the local
environment of each QD. These latter effects will not be considered
here.
\par The time evolution from any initial state under the
action of $H$ in Eq.(1) is
easily performed by means of the pseudo 1/2-spin operator formalism
\cite{qj,wokaun}.
The exact kinetic equations for this system can be obtained by 
applying the method of operator-equation hierarchy developed for
Dicke systems in
\cite{bogo}. As a test, we
verified that in the limit of zero laser intensity and no Forster term
our results coincide with those in \cite{palma} where two-state systems
coupled to a dephasing environment were considered. Until now the
experimental
identification and quantification of exciton decoherence mechanisms in
low dimensional
semiconductor heterostructures is rather scarce.
As a consequence we adopt here a simplified model. In a standard way, by assuming
a very short correlation time for exciton operators, the exact
hierarchy
of equations transforms into a Markovian master equation.  The initial
condition is represented by the density matrix
$\rho(0)=|0\rangle\langle 0|
\rho_{Ph}(T)$,
exciton vacuum and the equilibrium phonon reservoir
at temperature $T$.
At resonance, $i.e.$ $\Delta \omega=0$, the dynamical equation for the
expectation value of exciton operators is then given by
\begin{eqnarray}
\frac {\partial \langle J_{\alpha}^{r-s}\rangle}{\partial t}=
-iV\langle [J_{\alpha}^{r-s},J_z^2]\rangle
-iA\langle [J_{\alpha}^{r-s},J^++J^-]\rangle\\
\nonumber
-\Gamma(2\langle [J_{\alpha}^{r-s},J_z]J_z\rangle - 
\langle [J_{\alpha}^{r-s},J_z^2] \rangle)
\end{eqnarray}
where the decoherence rate is 
$\Gamma=\int {d\omega' \omega'^n
e^{-\omega'/\omega_c}(1+2N(\omega',T))}$
with $n$ depending on the dimensionality of the
phonon field, $\omega_c$ is a cut-off frequency (typically the Debye
frequency) and
$N(\omega',T)$ is the phonon Bose-Einstein occupation factor.
We do not attempt here to perform a microscopic calculation of $\Gamma$ but
instead we take it as a variable parameter.
We consider pure decoherence effects that do not involve energy
relaxation of
excitons, as indicated by the last term in Eq. (1).
\par It is a well known fact that
very narrow linewidth of the photoluminescence signal of
a {\it single} QD does exist due to the elimination of
inhomogeneous broadening effects. Consequently, 
the decoherence rate $\Gamma$ in our calculations should be associated with just
homogeneous broadening effects.
At low temperature the main decoherence
mechanism is indeed acoustic phonon scattering processes. The
decoherence parameter
$\Gamma$
is temperature dependent and it amounts for 20-50 $\mu$eV for typical
III-V
semiconductor QDs in a temperature range from 10 K to 30 K \cite{taka}.
We solve numerically 
the coupled differential linear equations for
the time dependent pseudo-spin expectation values (8 for Bell states and
15
for GHZ states). For $\Gamma$ we take typical values which can represent
real situations for QDs at low temperatures. Other common parameters for
the
results shown below are: resonance condition $\epsilon=\omega=1$ and
Forster term
$V=\epsilon/10$. Laser strengths and decoherence rates are to be
expressed in units
of $V$. As a quantitative measure of the successful generation of
exciton MES
we present our results in terms of the time dependent overlaps
$O_B(t)=Tr\{\rho_{Bell}\rho(t)\}$ and $O_G(t)=Tr\{\rho_{GHZ}\rho(t)\}$
where
$\rho_{Bell}=(1+J_z^{0-1}-J_z^{1-2})/3-J_y^{0-2}$ and
$\rho_{GHZ}=1/4+(J_z^{0-1}-J_z^{2-3})/2-J_y^{0-3}$ (we use the same
notation as in \cite{qj}).
$|0\rangle$ represents the exciton vacuum,
$|1\rangle$ denotes a single-exciton state,
$|2\rangle$ represents the biexciton state 
and $|3\rangle$ is the triexciton state.
\par In order to appreciate the importance of the non-linear Forster
term
to generate exciton MES we present in Fig. 1 the evolution of the
overlaps $O_B(t)$ and $O_G(t)$ in the
limit of very weak light excitation and zero decoherence \cite{qj}.
It is worth noting that no exciton MES generation is possible if the
Forster interaction is turned off. This implies that
efficient exciton MES generation should be helped by compact
QD systems where the Forster term can take a significant value.
\par Next, we discuss the $N=2$ case and Bell-state generation in
presence of noise. In Fig. 2a
results are shown for a decoherence rate $\Gamma=0.001$ and 
different laser intensities ($A=0.1$ and $A=0.4$). Bell-state generation
time is significantly shortened
by applying stronger laser pulses. Therefore, decoherence effects can be
minimized by using higher excitation levels. However, a higher laser
intensity
also implies a sharper evolution which therefore requires a very precise pulse
length.
In Fig. 3a Bell-state generation
is shown for different values of the decoherence parameter
($\Gamma=0.001, 0.01$ and $0.1$).
It is evident that at high temperature $\Gamma=0.1$ no MES generation is
possible.
However, we estimate that $\Gamma$ values between $0.001-0.01$ are
typical in the
temperature range from $10$ K to $50$ K. We conclude that a parameter
window exists 
where successful generation of Bell MES can be produced.
\par Now we address the GHZ MES generation in a $N=3$ QD system. As for
the Bell case, using higher laser excitation levels
it is possible to obtain in shorter times a total overlap with the GHZ
density matrix as depicted in Fig. 2b ($\Gamma=0.001$). Temperature effects 
through the variation of $\Gamma$
are depicted in Fig. 3b ($A=0.4$).  It is evident that similar decoherence rates yield 
a more dramatic reduction of the MES coherence in the GHZ case than in the Bell 
case. However, as for Bell generation, a parameter window
does exist where the generation of such entangled states can be
feasible.
\par It is worth noting the different scaling behaviour of the generation
frequency of these MES at very low temperature, $i.e.$ vanishing $\Gamma$ and
very low laser excitation. While 
selective $\pi/2$ laser pulse length for the Bell case
scales like $V/A^2$, selective $\pi/2$ pulse length for the
GHZ case scales like $V^2/A^3$. This property of $\pi/2$ pulses to generate
exciton MES was demonstrated in an analytical way in \cite{qj} and can be
verified in our numerical results by looking at Fig. 2a and Fig. 2b. 
\par In summary, we have shown that decoherence effects can be minimized
in
the generation of maximally entangled states by applying stronger laser
pulses and working at low temperatures where acoustic phonon scattering
is
the main decoherence mechanism.

\par This work has been partially supported by COLCIENCIAS.

\newpage

\centerline{\bf Figure Captions}

\bigskip

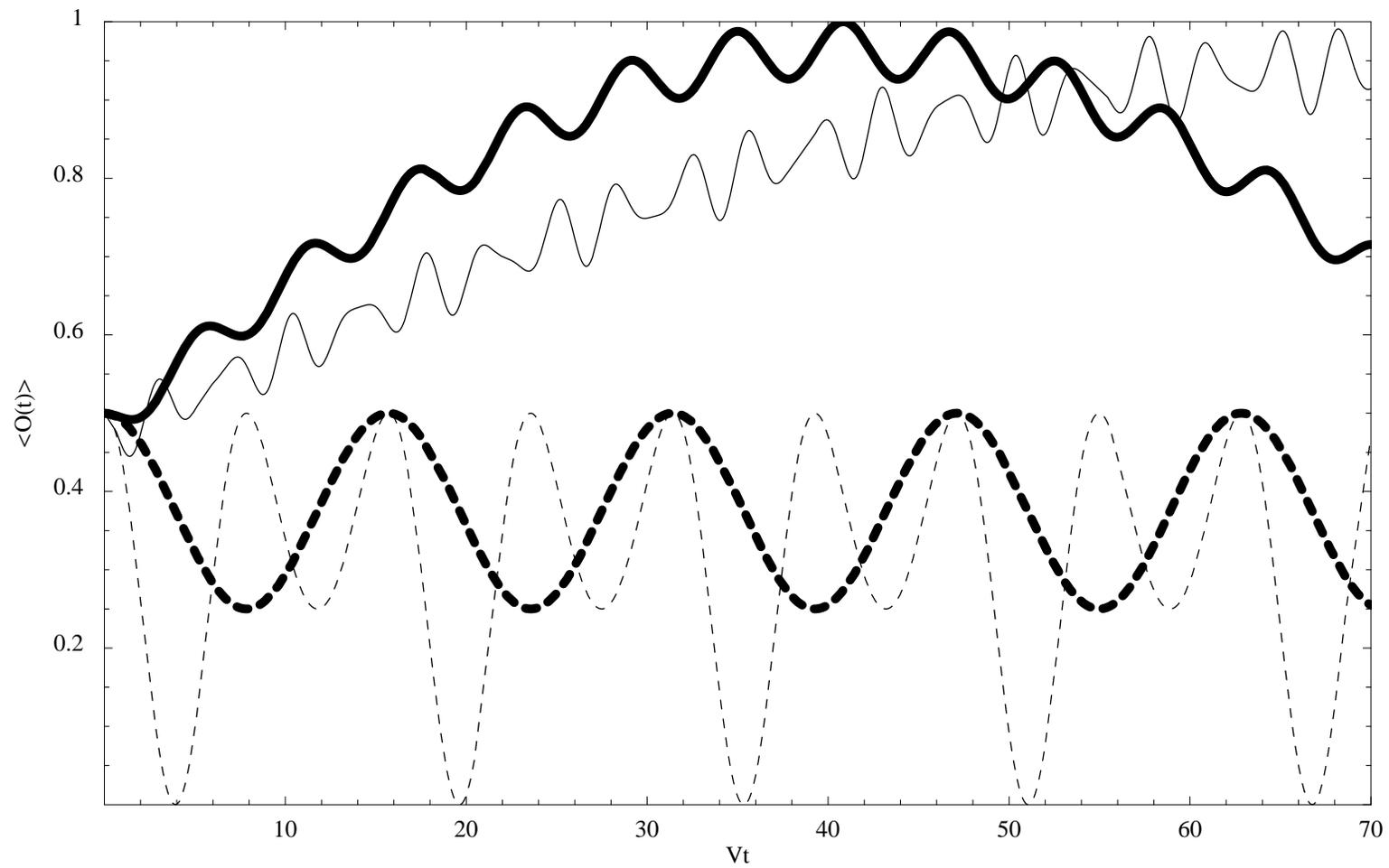
\noindent Figure 1: Exciton MES generation in the zero decoherence limit. Thick 
lines represents the 
Bell-state overlap with $A=0.1$: solid, Forster term included; dotted, 
Forster term not included. Thin lines represent the GHZ-state overlap with 
$A=0.2$ and similar meaning for solid and dotted lines.

\bigskip

\noindent Figure 2: Exciton MES generation in the presence of decoherence
 (a)  $\langle O_B(t)\rangle$ for $A=0.1$, dotted line and
$A=0.4$, solid line. (b) $\langle O_G(t) \rangle$ for $A=0.2$, dotted line and
$A=0.4$, solid line. $\Gamma=0.001$. 

\bigskip

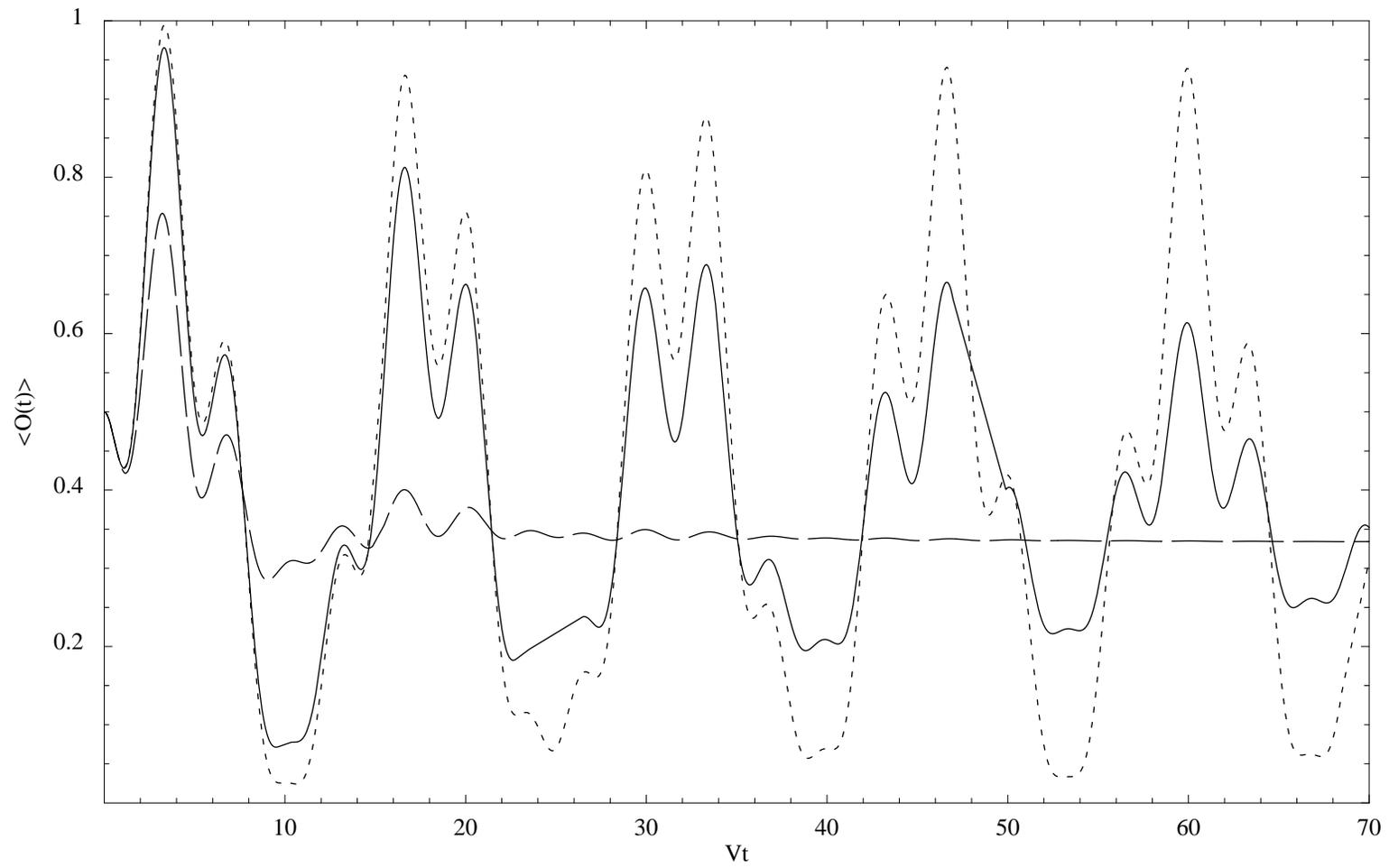
\noindent Figure 3: Exciton MES generation in the presence of decoherence
(a) $\langle O_B(t)\rangle$ and
(b) $\langle O_G(t) \rangle$. $A=0.4$, $\Gamma=0.001$, dotted line,
$\Gamma=0.01$, solid line and $\Gamma=0.1$, dashed line for both (a) and (b).

\end{document}